\PassOptionsToPackage{unicode}{hyperref}
\PassOptionsToPackage{hyphens}{url}
\documentclass[
]{article}
\usepackage{amsmath,amssymb}
\usepackage{lmodern}
\usepackage{iftex}
\ifPDFTeX
  \usepackage[T1]{fontenc}
  \usepackage[utf8]{inputenc}
  \usepackage{textcomp} 
\else 
  \usepackage{unicode-math}
  \defaultfontfeatures{Scale=MatchLowercase}
  \defaultfontfeatures[\rmfamily]{Ligatures=TeX,Scale=1}
\fi
\IfFileExists{upquote.sty}{\usepackage{upquote}}{}
\IfFileExists{microtype.sty}{
  \usepackage[]{microtype}
  \UseMicrotypeSet[protrusion]{basicmath} 
}{}
\makeatletter
\@ifundefined{KOMAClassName}{
  \IfFileExists{parskip.sty}{%
    \usepackage{parskip}
  }{
    \setlength{\parindent}{0pt}
    \setlength{\parskip}{6pt plus 2pt minus 1pt}}
}{
  \KOMAoptions{parskip=half}}
\makeatother
\usepackage{xcolor}
\IfFileExists{xurl.sty}{\usepackage{xurl}}{} 
\IfFileExists{bookmark.sty}{\usepackage{bookmark}}{\usepackage{hyperref}}
\hypersetup{
  hidelinks,
  pdfcreator={LaTeX via pandoc}}
\usepackage{longtable,booktabs,array}
\usepackage{calc} 
\usepackage{etoolbox}
\makeatletter
\patchcmd\longtable{\par}{\if@noskipsec\mbox{}\fi\par}{}{}
\makeatother
\IfFileExists{footnotehyper.sty}{\usepackage{footnotehyper}}{\usepackage{footnote}}
\makesavenoteenv{longtable}
\usepackage{graphicx}
\makeatletter
\def\maxwidth{\ifdim\Gin@nat@width>\linewidth\linewidth\else\Gin@nat@width\fi}
\def\maxheight{\ifdim\Gin@nat@height>\textheight\textheight\else\Gin@nat@height\fi}
\makeatother
\setkeys{Gin}{width=\maxwidth,height=\maxheight,keepaspectratio}
\makeatletter
\def\fps@figure{htbp}
\makeatother
\providecommand{\tightlist}{%
  \setlength{\itemsep}{0pt}\setlength{\parskip}{0pt}}
\newlength{\cslhangindent}
\newlength{\csllabelwidth}
\newlength{\cslentryspacingunit} 
\newenvironment{CSLReferences}[2] 
 {
  \setlength{\parindent}{0pt}
  \ifodd #1
  \let\oldpar\par
  \def\par{\hangindent=\cslhangindent\oldpar}
  \fi
  \setlength{\parskip}{#2\cslentryspacingunit}
 }%
 {}
\usepackage{calc}

\newcommand{\CSLLeftMargin}[1]{\parbox[t]{\csllabelwidth}{#1}}
\newcommand{\CSLRightInline}[1]{\parbox[t]{\linewidth - \csllabelwidth}{#1}\break}

\ifPDFTeX
  \DeclareUnicodeCharacter{2032}{\ensuremath{^{\prime}}}
  \DeclareUnicodeCharacter{2033}{\ensuremath{^{\prime\prime}}}
  \DeclareUnicodeCharacter{2070}{\ensuremath{^{0}}}
  \DeclareUnicodeCharacter{2074}{\ensuremath{^{4}}}
  \DeclareUnicodeCharacter{2075}{\ensuremath{^{5}}}
  \DeclareUnicodeCharacter{2076}{\ensuremath{^{6}}}
  \DeclareUnicodeCharacter{2077}{\ensuremath{^{7}}}
  \DeclareUnicodeCharacter{2078}{\ensuremath{^{8}}}
  \DeclareUnicodeCharacter{2079}{\ensuremath{^{9}}}
  \DeclareUnicodeCharacter{2194}{\ensuremath{\leftrightarrow}}
  \DeclareUnicodeCharacter{2200}{\ensuremath{\forall}}
  \DeclareUnicodeCharacter{2202}{\ensuremath{\partial}}
  \DeclareUnicodeCharacter{2203}{\ensuremath{\exists}}
  \DeclareUnicodeCharacter{2205}{\ensuremath{\emptyset}}
  \DeclareUnicodeCharacter{2207}{\ensuremath{\nabla}}
  \DeclareUnicodeCharacter{2208}{\ensuremath{\in}}
  \DeclareUnicodeCharacter{2209}{\ensuremath{\notin}}
  \DeclareUnicodeCharacter{2211}{\ensuremath{\sum}}
  \DeclareUnicodeCharacter{2212}{\ensuremath{-}}
  \DeclareUnicodeCharacter{2213}{\ensuremath{\mp}}
  \DeclareUnicodeCharacter{2227}{\ensuremath{\land}}
  \DeclareUnicodeCharacter{2228}{\ensuremath{\lor}}
  \DeclareUnicodeCharacter{2229}{\ensuremath{\cap}}
  \DeclareUnicodeCharacter{2243}{\ensuremath{\simeq}}
  \DeclareUnicodeCharacter{2245}{\ensuremath{\cong}}
  \DeclareUnicodeCharacter{2248}{\ensuremath{\approx}}
  \DeclareUnicodeCharacter{2260}{\ensuremath{\neq}}
  \DeclareUnicodeCharacter{2261}{\ensuremath{\equiv}}
  \DeclareUnicodeCharacter{2264}{\ensuremath{\leq}}
  \DeclareUnicodeCharacter{2265}{\ensuremath{\geq}}
  \DeclareUnicodeCharacter{2282}{\ensuremath{\subset}}
  \DeclareUnicodeCharacter{2286}{\ensuremath{\subseteq}}
  \DeclareUnicodeCharacter{223C}{\ensuremath{\sim}}
  \DeclareUnicodeCharacter{221D}{\ensuremath{\propto}}
  \DeclareUnicodeCharacter{221E}{\ensuremath{\infty}}
  \DeclareUnicodeCharacter{221A}{\ensuremath{\surd}}
  \DeclareUnicodeCharacter{220F}{\ensuremath{\prod}}
  \DeclareUnicodeCharacter{222B}{\ensuremath{\int}}
  \DeclareUnicodeCharacter{222A}{\ensuremath{\cup}}
  \DeclareUnicodeCharacter{22C5}{\ensuremath{\cdot}}
  \DeclareUnicodeCharacter{21D2}{\ensuremath{\Rightarrow}}
  \DeclareUnicodeCharacter{21D4}{\ensuremath{\Leftrightarrow}}
  \DeclareUnicodeCharacter{207A}{\ensuremath{^{+}}}
  \DeclareUnicodeCharacter{207B}{\ensuremath{^{-}}}
  \DeclareUnicodeCharacter{0393}{\ensuremath{\Gamma}}
  \DeclareUnicodeCharacter{0394}{\ensuremath{\Delta}}
  \DeclareUnicodeCharacter{0398}{\ensuremath{\Theta}}
  \DeclareUnicodeCharacter{039B}{\ensuremath{\Lambda}}
  \DeclareUnicodeCharacter{039E}{\ensuremath{\Xi}}
  \DeclareUnicodeCharacter{03A0}{\ensuremath{\Pi}}
  \DeclareUnicodeCharacter{03A3}{\ensuremath{\Sigma}}
  \DeclareUnicodeCharacter{03A6}{\ensuremath{\Phi}}
  \DeclareUnicodeCharacter{03A8}{\ensuremath{\Psi}}
  \DeclareUnicodeCharacter{03A9}{\ensuremath{\Omega}}
  \DeclareUnicodeCharacter{03B1}{\ensuremath{\alpha}}
  \DeclareUnicodeCharacter{03B2}{\ensuremath{\beta}}
  \DeclareUnicodeCharacter{03B3}{\ensuremath{\gamma}}
  \DeclareUnicodeCharacter{03B4}{\ensuremath{\delta}}
  \DeclareUnicodeCharacter{03B5}{\ensuremath{\varepsilon}}
  \DeclareUnicodeCharacter{03B6}{\ensuremath{\zeta}}
  \DeclareUnicodeCharacter{03B7}{\ensuremath{\eta}}
  \DeclareUnicodeCharacter{03B8}{\ensuremath{\theta}}
  \DeclareUnicodeCharacter{03B9}{\ensuremath{\iota}}
  \DeclareUnicodeCharacter{03BA}{\ensuremath{\kappa}}
  \DeclareUnicodeCharacter{03BB}{\ensuremath{\lambda}}
  \DeclareUnicodeCharacter{03BC}{\ensuremath{\mu}}
  \DeclareUnicodeCharacter{03BD}{\ensuremath{\nu}}
  \DeclareUnicodeCharacter{03BE}{\ensuremath{\xi}}
  \DeclareUnicodeCharacter{03C0}{\ensuremath{\pi}}
  \DeclareUnicodeCharacter{03C1}{\ensuremath{\rho}}
  \DeclareUnicodeCharacter{03C2}{\ensuremath{\varsigma}}
  \DeclareUnicodeCharacter{03C3}{\ensuremath{\sigma}}
  \DeclareUnicodeCharacter{03C4}{\ensuremath{\tau}}
  \DeclareUnicodeCharacter{03C5}{\ensuremath{\upsilon}}
  \DeclareUnicodeCharacter{03C6}{\ensuremath{\phi}}
  \DeclareUnicodeCharacter{03C7}{\ensuremath{\chi}}
  \DeclareUnicodeCharacter{03C8}{\ensuremath{\psi}}
  \DeclareUnicodeCharacter{03C9}{\ensuremath{\omega}}
\fi
\usepackage{orcidlink}
\ifLuaTeX
  \usepackage{selnolig}  
\fi

\author{}
\date{}

\begin{document}

\hypertarget{re-randomization-probability-weighting-for-population-adjusted-comparisons-of-maintenance-outcomes-across-trial-designs}{%
\section{Re-randomization-probability weighting for population-adjusted
comparisons of maintenance outcomes across trial
designs}\label{re-randomization-probability-weighting-for-population-adjusted-comparisons-of-maintenance-outcomes-across-trial-designs}}

\textbf{Kim Rand, PhD}~\orcidlink{0000-0001-7692-4099}

Health Services Research Centre, Akershus University Hospital,
Lørenskog, Norway

Maths in Health B.V., Klimmen, the Netherlands

\hypertarget{abstract}{%
\subsection{Abstract}\label{abstract}}

Randomized-withdrawal trials re-randomize induction responders to
continued treatment or withdrawal. Baseline characteristics are often
reported for the induction population but unavailable for responders,
limiting population-adjusted indirect comparisons of maintenance
outcomes. We describe re-randomization-probability weighting, which
combines induction-population calibration with inverse weighting by
known maintenance assignment probabilities. Coding induction
non-response as failure defines durable response: response at both
induction and maintenance. The design weights reproduce
induction-population covariate totals in randomization expectation,
allowing population adjustment using reported induction baselines.
Conditional maintenance probabilities can also be estimated, although
different induction treatments generate different responder populations.
A synthetic three-trial demonstration reduced durable-response bias from
0.106 to 0.001 and from 0.055 to 0.001 in two comparisons.
Full-procedure simulations showed interval coverage of 91.5\%--94.3\%;
stress scenarios illustrated bias from omitted covariates and incorrect
assignment weights. The method connects established design weighting and
aggregate dilution approaches to population-adjusted comparison across
randomized-withdrawal and treat-through designs. It requires compatible
durable-response endpoints and credible transport and modeling
assumptions. Withdrawal after different induction regimens is not
automatically a common comparator. Where no valid anchor exists,
unanchored comparison is appropriate only if its demanding adjustment
assumptions are defensible.

\textbf{Keywords:} matching-adjusted indirect comparison;
population-adjusted indirect comparison; randomized withdrawal;
re-randomization; maintenance of response; evidence synthesis

\hypertarget{highlights}{%
\subsection{Highlights}\label{highlights}}

\textbf{What is already known:} Randomized-withdrawal (responder
re-randomization) trials inconsistently report baseline characteristics
for the re-randomized subset, complicating population-adjusted indirect
comparison of maintenance outcomes. In network meta-analysis,
re-randomized trials have been converted to approximate treat-through
arms by diluting induction non-responders by the re-randomization
factor.

\textbf{What is new:} This paper shows that calibrating the full
induction cohort to the comparator's reported induction-level aggregates
and weighting maintenance outcomes by the known re-randomization
probabilities yields induction-population estimands of durable response.
The same design identity, applied as weights, in expectation
reconstructs the induction population's baseline covariate distribution
within each maintenance arm, for any function of baseline covariates.
The approach can be used to harmonize trials with different
re-randomization ratios and treat-through designs. The paper also shows
that withdrawal arms are not a common anchor for durable response across
different induction regimens, and that unanchored comparison is
defensible only under its full adjustment assumptions.

\textbf{Potential impact:} Analysts conducting indirect comparisons of
maintenance outcomes, and health technology assessment bodies reviewing
them, gain a simple, transparent method applicable where required
baseline data for standard approaches do not exist.

\hypertarget{background}{%
\subsection{Background}\label{background}}

In the absence of head-to-head randomized controlled trials (RCTs),
indirect treatment comparisons (ITCs) are used to estimate comparative
effectiveness between interventions. Naive (direct) comparison of
absolute outcomes from separate trials may easily be biased by
differences in the distribution of baseline characteristics between
different trial populations. Consequently, population-adjusted indirect
comparison (PAIC) methods such as matching-adjusted indirect comparison
(MAIC)\textsuperscript{1} and simulated treatment comparison
(STC)\textsuperscript{2} have become increasingly common in submissions
to health technology assessment (HTA) bodies such as the National
Institute for Health and Care Excellence (NICE) in the
UK\textsuperscript{3}. The NICE Decision Support Unit has provided
methodological guidance on use of PAIC methods in appraisal
submissions\textsuperscript{4}. PAIC aims to reduce bias in indirect
comparisons by adjusting for differences in effect modifiers and, in
unanchored comparisons, prognostic variables between trial populations,
thereby approximating the conditions of a valid treatment comparison.
PAIC methods rely on access to individual-level patient data (IPD) from
at least one index clinical trial, as well as adequately reported
marginal statistics for all relevant baseline characteristics from the
comparator trial or trials.

Briefly, MAIC involves re-weighting the IPD from the index trial such
that the weighted marginal statistics (means and, where reported,
standard deviations) of baseline covariates align with the reported
aggregate-level characteristics in the comparator
trial\textsuperscript{5}. STC employs regression modeling to estimate
outcomes as a function of baseline covariates. The model is then
standardized to the comparator trial's reported population. In both
cases, the validity of the adjustment depends on the availability and
quality of reported baseline characteristics in the comparator trial. In
cases where effect modifiers or prognostic factors are missing or
inadequately handled, the bias may persist or even be
exacerbated\textsuperscript{4}.

Inadequate reporting of baseline characteristics is particularly acute
in clinical trials using randomized withdrawal or randomized
discontinuation designs. In these trials, patients achieving treatment
specific response criteria following administration of active treatment
in the induction period are re-randomized to continued treatment or
treatment withdrawal/placebo for the maintenance period. This contrasts
with treat-through designs, in which all patients continue their
randomized treatment regardless of induction response. Responder
re-randomization designs are used in several clinical areas to evaluate
maintenance of treatment effect and to inform long-term benefit-risk
assessment, but they introduce complexities for indirect comparisons.
Consider that patients who achieve clinical response are selected on a
post-baseline outcome: the responder subset will typically differ
systematically from the initial trial population in its baseline
characteristics. Although baseline characteristics are generally
well-reported for the initial trial population, the corresponding
characteristics of the re-randomized responders are often sparsely
reported or entirely unavailable. Reporting practice varies even within
disease area. For instance, among recent randomized-withdrawal trials of
biologics in atopic dermatitis, the ECZTRA 1 and 2 publication reports
baseline characteristics for the randomized induction population
only\textsuperscript{6}, the SOLO-CONTINUE publication describes the
re-randomized arms at the point of re-randomization rather than at
pre-treatment baseline\textsuperscript{7}, and only the ADvocate
maintenance publication tabulates week-0 characteristics for the
re-randomized subset\textsuperscript{8}. Analysts cannot rely on this
availability. Standard PAIC methods are therefore difficult to apply to
maintenance outcomes: matching on the responder population is not
possible in the absence of reported baseline characteristics, and
matching on the reported induction-population baselines is incorrect for
a responder-level comparison, since the responder subset differs
systematically from the induction population.

One piece of the required information is ordinarily published even when
responder baselines are not: for fixed-ratio designs, the
re-randomization ratio is part of the design description. (Stratified or
adaptive designs additionally require stratum-specific assignment
detail, which may or may not be reported; we return to this in the
Methods.) Such information on re-randomization ratios has been exploited
in network meta-analyses (NMAs) at the level of aggregate outcome
counts. When combining trials of different designs, re-randomized
maintenance data have been converted to counterfactual treat-through
arms by combining the responders assigned to a given maintenance arm
with the induction non-responders divided (``diluted'') by the
re-randomization factor\textsuperscript{9}. This conversion has since
been applied in several NMAs of maintenance
treatment\textsuperscript{10--12}, and has been carried into health
technology appraisals, where evidence reviewers describe its key
weakness explicitly: the converted comparison ``assumes that there are
no systematic differences between the baseline characteristics of
induction responders and non-responders''\textsuperscript{13}. This
``dilution'' approach has also been run in the opposite direction:
treat-through trials have been adjusted to mimic re-randomized designs
by restriction to their induction-responder subsets, with the
induction-response-times-maintenance composition then applied at the
treatment level to simulate intent-to-treat efficacy of the sequential
regimens\textsuperscript{14}. Others have judged the two designs not
comparable at all and analysed them in separate networks, leaving
treatments studied under different designs
uncompared\textsuperscript{15}; current practice thus spans
outcome-level conversion on implicit assumptions at one end and refusal
to compare at the other.

Separately, recent MAICs involving responder re-randomization have
matched on responder-population or re-randomization-timepoint
characteristics where these happened to be
reported\textsuperscript{16--18} (with Augustin et al.~using both
induction baseline and maintenance baseline characteristics in their
matching), or reconstructed responder-population baseline summaries from
induction-cohort aggregates where they were not\textsuperscript{19}.
However, the literature does not point out that the same justification
behind the outcome dilution approach also produces, in expectation, the
baseline covariate distribution of the induction population for each
maintenance arm. This allows for population adjustment at the induction
level when responder-level baseline characteristics are unavailable. Our
team has used this reconstruction method in several ITC projects, the
first of which has recently been publicly disclosed, including a brief
methods description\textsuperscript{20}. A formal account of the general
method's form, conditions, and limitations has, to our knowledge, not
been given.

This paper provides a detailed account of the method, which can be
construed as a generalization of the dilution approach, from aggregate
outcome counts to arbitrary functions of baseline covariates, directly
usable with IPD. RPW uses the same fractional allocation of induction
non-responders as the earlier dilution conversions; here, however, their
outcome contribution is fixed at zero because the endpoint requires
response at both assessments, whereas conversions that include late
responders target a different outcome. The procedure combines
calibration of the full induction cohort to match the comparator's
reported induction aggregates with inverse-probability weighting using
known re-randomization probabilities. We refer to the design-weight
approach as \emph{re-randomization-probability weighting} (RPW). RPW
enables PAIC for binary maintenance outcomes adjusted to the
induction-population level, requiring that baseline characteristics of
the full induction population are adequately reported for the comparator
in question and that the re-randomization probabilities have been
adequately described. Because the reconstruction approach works with any
randomized-withdrawal design, RPW can also be used to harmonize trials
with different re-randomization ratios, as well as randomized-withdrawal
trials with treat-through trials. This paper presents the method,
provides a fully reproducible demonstration case in R, and discusses
assumptions, strengths, and limitations. We also examine the merits and
requirements for using the withdrawal arms as a shared reference in
anchored PAIC.

\hypertarget{methods}{%
\subsection{Methods}\label{methods}}

\hypertarget{setting-and-notation}{%
\subsubsection{Setting and notation}\label{setting-and-notation}}

Consider an active-treatment cohort in a randomized withdrawal trial.
Let \(X_i\) denote the vector of baseline (pre-treatment) covariates for
patient \(i = 1, \ldots, n\), and let \(R_i \in \{0, 1\}\) indicate
achievement of protocol-defined response by the end of induction.
Responders (\(R_i = 1\)) are re-randomized to one of \(K\) maintenance
arms; let \(Z_i \in \{1, \ldots, K\}\) denote the maintenance
assignment, with probabilities
\(\pi_{ik} = \Pr(Z_i = k \mid R_i = 1, S_i)\), where \(S_i\) contains
any strata used in the re-randomization. With non-stratified
re-randomization in a fixed ratio, \(\pi_{ik} \equiv \pi_k\) for all
\(i\). Non-responders are typically not followed beyond induction, or
are followed separately from responders. Let \(M_i \in \{0, 1\}\)
indicate response following maintenance, observed only for re-randomized
responders. Finally, define durable response \(D_i = R_i M_i\), with the
convention that \(R_i = 0 \Rightarrow D_i = 0\): a patient who does not
achieve response following induction has no response to maintain at the
end of maintenance.

Response following induction is inherently treatment-dependent: it
depends on the induction regimen received. This dependency is suppressed
above for notational convenience, but it becomes central when we discuss
anchoring below. We therefore write \(R^{(j)}\) for response following
induction regimen \(j\), and \(M^{(j,k)}\) for response during
maintenance when induction regimen \(j\) is followed by maintenance
regimen \(k\). An important consequence of this notation is that durable
response is a property of a sequential regimen (induction \(j\), then
maintenance \(k\)), not of the maintenance treatment alone.

\hypertarget{re-randomization-probability-weighting}{%
\subsubsection{Re-randomization-probability
weighting}\label{re-randomization-probability-weighting}}

The general RPW construction is an inverse-probability weight applied to
the whole induction cohort. For maintenance arm \(k\), let

\[W_{ik} = (1 - R_i) + R_i \, \frac{\mathbf{1}(Z_i = k)}{\pi_{ik}}.\]

Under valid re-randomization,
\(E[\mathbf{1}(Z_i = k) \mid R_i = 1, X_i, S_i] = \pi_{ik}\), and
therefore

\[E[W_{ik} \mid X_i, R_i, S_i] = 1,\]

so that, for any integrable function \(h\) of baseline covariates and
known \(\pi_{ik} > 0\),

\[E\left[\frac{1}{n} \sum_{i=1}^{n} W_{ik} \, h(X_i) \,\middle|\, X, R, S\right] = \frac{1}{n} \sum_{i=1}^{n} h(X_i).\]

That is, conditional on the realized induction cohort \((X, R, S)\)
(covariates and induction responses for all patients, and strata among
responders), the weighted pseudo-population of arm \(k\) reproduces in
expectation the empirical baseline covariate distribution of the total
induction population. This is the familiar inverse-probability-weighting
representation used in sequentially randomized
designs\textsuperscript{21,22}, and it holds for stratified and
otherwise unequal assignment probabilities alike. Non-responders always
receive weight 1, so no assignment probability needs to be defined for
them. This is convenient when re-randomization strata are defined by
characteristics assessed after induction, which may not be defined for
non-responders.

With a fixed re-randomization ratio, \(\pi_{ik} \equiv \pi_k\) for all
patients, rescaling by the common factor gives the equivalent design
weight

\[\omega_{ik} = \pi_k W_{ik} = R_i \, \mathbf{1}(Z_i = k) + (1 - R_i) \, \pi_k.\]

Responders re-randomized to arm \(k\) receive weight 1, responders
re-randomized to any other arm receive weight 0, and induction
non-responders receive the fractional weight \(\pi_k\). This
fractional-weight form applies only when the ratio is fixed across
patients. Substituting patient-specific \(\pi_{ik}\) for \(\pi_k\) in
\(\omega_{ik}\) does not reconstruct the induction population. The
weighted distribution then targets the \(\pi_{ik}\)-weighted covariate
distribution, \(\sum_i \pi_{ik} h(X_i) / \sum_i \pi_{ik}\), which
differs from the unweighted one when assignment probabilities vary
systematically, for example by strata or with baseline characteristics.
What matters is variation between patients. An unequal ratio that is the
same for every patient (e.g.~2:2:1 in a three-arm design) is still a
fixed-ratio case, whereas assignment probabilities that vary between
patients or strata require the \(W_{ik}\) form. Stratification does not
by itself imply varying assignment probabilities, because the same ratio
can be applied within every stratum. Stratified re-randomization is not
uncommon; in SOLO-CONTINUE, for example, induction responders were
re-randomized in a 2:1:1:1 ratio stratified by prior regimen, region,
and current disease severity\textsuperscript{7}.

Informally, in the fixed-ratio case, a patient entering induction
becomes either a responder or a non-responder. A responder is assigned
to maintenance arm \(k\) with probability \(\pi_k\), and a non-responder
is included in the arm-\(k\) pseudo-population with fractional weight
\(\pi_k\). Either way, the patient's expected participation in the
arm-\(k\) pseudo-population is \(\pi_k\), so the pseudo-population is an
unbiased weighted sample of the induction population, whose baseline
characteristics are typically well reported.

``Unbiased'' needs to be stated carefully. Linear functionals of the
empirical measure (e.g.~means, proportions, raw moments) are reproduced
exactly in expectation, in the manner of Horvitz--Thompson
estimation\textsuperscript{23}. A normalized pseudo-arm summary such as
\(\sum_i \omega_{ik} h(X_i) / \sum_i \omega_{ik}\) has a random
denominator. It is a Hájek-type estimator: generally consistent and
often more stable, but not exactly unbiased in finite samples. Likewise,
non-linear functions of several moments or characteristics, such as the
variance, are reconstructed consistently but not unbiasedly. None of
this affects the practical use of the reconstruction as a calibration
target, but claims of exact unbiasedness are limited to linear
functionals.

A related practical choice is between design-specified and realized
assignment. The realized allocation in a trial may deviate from the
design ratio: 1:1:1 re-randomization may, for instance, realize as 35\%,
32\%, and 33\% of responders. With fixed allocation ratios,
non-responder weights can be based on the realized proportions instead
of the design proportions. Conditional on the realized arm sizes, the
responders assigned to arm \(k\) form a random sample of all responders.
Using realized proportions therefore reproduces the induction-population
covariate distribution exactly in conditional expectation and removes
the random variation in the number of responders assigned to each arm.
Formally, this is the Hájek-type counterpart of the design weight, and
it corresponds to the fully conditioned product form described under
Estimation with population adjustment below. I use the design-specified
\(\pi_{ik}\) for the primary RPW definition for two reasons. First, it
gives the exact identity in expectation over the randomization described
above. Second, it extends directly to inverse-probability weighting with
unequal assignment, for example under stratification, where realized
proportions within strata may be unstable or even undefined.

\hypertarget{estimands}{%
\subsubsection{Estimands}\label{estimands}}

Coding non-responders as maintenance failures makes the durable-response
outcome \(D_i = R_i M_i\) well defined whether or not non-responders are
followed up. Let \(r_T = \Pr_T(R = 1)\) denote the induction response
probability in a target population \(T\). Two estimands are relevant in
\(T\):

\[q_{k,T} = \Pr_T\{R = 1, M(k) = 1\}, \qquad m_{k,T} = \Pr_T\{M(k) = 1 \mid R = 1\} = \frac{q_{k,T}}{r_T}.\]

The durable-response probability \(q_{k,T}\) is the probability of
responding at induction and then, under maintenance \(k\), at the
maintenance assessment. It is defined on the full induction population
and therefore does not condition on post-baseline events. The
conditional maintenance probability \(m_{k,T}\) is the probability of
maintaining response under \(k\) among those who responded following
induction. It can be calculated as the ratio of the standardized
durable-response and induction-response probabilities. Both are
potentially useful and both are reported in applied work; we return to
their relative merits in the Discussion. The general caution is that
\(m_{k,T}\) conditions on a post-baseline event. When trials administer
different induction treatments, or use different response definitions or
induction durations, their responder subsets are different selected
populations, and standardization to the same pre-treatment baseline does
not make them equivalent. A between-regimen difference in conditional
maintenance probabilities is therefore a contrast of regimen-specific
responder outcomes, not a causal effect of the maintenance treatments in
a common set of patients.

\(T\) denotes the target population to which estimates apply. As is
conventional for MAIC, \(T\) is typically the comparator trial's
population\textsuperscript{4}. For cross-trial comparisons of different
drugs the regimen index is restored:
\(q_{jk,T} = \Pr_T\{R^{(j)} = 1, M^{(j,k)} = 1\}\),
\(r_{j,T} = \Pr_T\{R^{(j)} = 1\}\), and
\(m_{jk,T} = q_{jk,T} / r_{j,T}\). This makes two facts explicit: the
responder population depends on the induction treatment, and the
durable-response probability is a property of the joint
induction-plus-maintenance regimen rather than of the maintenance
treatment alone. Within each trial, the durable-response estimand is
identified by the randomized design. All patients are followed through
induction, so induction response \(R_i\) is observed for everyone, and
responders are re-randomized, so maintenance response can be estimated
for each arm. Indirect comparison additionally requires the standard
PAIC transportability, overlap\textsuperscript{4} and adjustment
assumptions\textsuperscript{4}.

\hypertarget{identification-and-assumptions}{%
\subsubsection{Identification and
assumptions}\label{identification-and-assumptions}}

Three tiers of conditions may be distinguished.

\begin{enumerate}
\def\labelenumi{\arabic{enumi}.}
\tightlist
\item
  \textbf{Within-trial design identification.} Each trial identifies its
  own sequential-regimen quantities \(r\), \(q_k\) and \(m_k\) in its
  induction population under the standard design requirements:
  consistency and no interference; known, positive maintenance
  assignment probabilities \(\pi_{ik}\); valid re-randomization
  conditional on the strata \(S_i\); handling of missing maintenance
  outcomes compatible with the endpoint definition; and a well-defined
  binary composite durable response \(D_i = R_i M_i\). This is where the
  RPW design identity is used: it justifies calibration to the induction
  population when the outcome is a maintenance outcome.
\item
  \textbf{Cross-trial transport.} An explicitly named target
  population\textsuperscript{4}; measurement of all effect modifiers
  and, for unanchored comparison of absolute outcomes, all prognostic
  factors\textsuperscript{4}; positivity and overlap between the index
  and target populations\textsuperscript{4}; and compatible definitions
  of the endpoint, induction response, eligibility, rescue, follow-up
  and other intercurrent events across trials.
\item
  \textbf{Estimation and modeling.} Moment calibration balances only the
  selected reported moments, not the joint covariate distribution, so
  comparability must be assumed for any relevant features that depend on
  the joint distribution. Regression-based standardization additionally
  requires a correctly specified outcome model and adequate information
  about the target covariate distribution.
\end{enumerate}

The RPW design identity does not relax or weaken any of the conditions
in the second or third tier.

\hypertarget{estimation-with-population-adjustment}{%
\subsubsection{Estimation with population
adjustment}\label{estimation-with-population-adjustment}}

Let \(a_i\) denote a PAIC calibration weight estimated on the full
induction population. For MAIC, the weights are chosen so that the index
trial's weighted baseline moments match the marginal statistics reported
for the comparator trial's induction-population
baseline\textsuperscript{5}. This is the step RPW makes possible where
induction-level marginal statistics are reported. Estimators
standardized to the comparator population are then

\[\hat{r}_T = \frac{\sum_i a_i R_i}{\sum_i a_i}, \qquad \hat{q}_{k,T} = \frac{\sum_i a_i R_i \mathbf{1}(Z_i = k) M_i / \pi_{ik}}{\sum_i a_i}, \qquad \hat{m}_{k,T} = \frac{\sum_i a_i R_i \mathbf{1}(Z_i = k) M_i / \pi_{ik}}{\sum_i a_i R_i \mathbf{1}(Z_i = k) / \pi_{ik}}.\]

The estimator \(\hat{q}_{k,T}\) has the fixed normalization
\(\sum_i a_i\) and, in the manner of Horvitz--Thompson
estimation\textsuperscript{23}, is design-unbiased for the
\(a\)-weighted durable-response rate given the calibration weights.
Induction non-responders do not appear in its numerator, since they
cannot achieve durable response. The role of the reconstruction is to
justify the calibration itself. Three weight objects should be held
distinct: the calibration weights \(a_i\), which define the target
population; the maintenance-randomization weights \(1/\pi_{ik}\), which
enter the outcome estimators; and the pseudo-arm design weights
\(W_{ik}\) or \(\omega_{ik}\), which justify the reconstruction and
serve as a proof device and balance diagnostic. Calibration weights are
estimated once, on the full induction cohort. Design weights enter
outcome estimation and pseudo-arm diagnostics, and separate calibration
of the pseudo-arms is not required.

For the conditional maintenance probability, \(\hat{m}_{k,T}\) takes the
Hájek form shown above. Under a fixed ratio the \(1/\pi_k\) factors
cancel, and it reduces to the calibration-weighted maintenance success
rate among responders assigned to arm \(k\). This is the form used in
the demonstration. The alternative ratio \(\hat{q}_{k,T}/\hat{r}_T\) is
consistent for the same estimand but noisier in finite samples. Its
numerator carries re-randomization count noise (the realized number of
responders assigned to arm \(k\)), which does not cancel against the
denominator. In the demonstration below, its 95\% percentile interval is
about one and a half times as wide as that of the direct form (0.33
versus 0.22). The same distinction arises for \(q\) itself, and three
finite-sample forms should be held apart:

\begin{itemize}
\tightlist
\item
  \textbf{Horvitz--Thompson form:} \(\hat{q}_{k,T}\) as defined above,
  with the fixed denominator \(\sum_i a_i\). It is design-unbiased given
  the calibration weights, but finite-sample estimates can fall outside
  the probability bounds, so the form cannot be logit-transformed
  without restriction.
\item
  \textbf{Ratio-normalized pseudo-arm form:}
  \(\sum_i a_i \omega_{ik} D_i / \sum_i a_i \omega_{ik}\), which
  normalizes by the realized pseudo-arm weight total while retaining the
  design-specified \(\pi_k\) in the weights. It is consistent for the
  same estimand and respects the probability bounds. The normalization
  reduces, but does not fully remove, the variation arising from
  realized allocation counts, and any precision gain is common but not
  guaranteed across outcomes and designs.
\item
  \textbf{Fully conditioned product form:} replacing the
  design-specified \(\pi_k\) with the realized (calibration-weighted)
  assignment proportions yields exactly
  \(\hat{r}_T \times \hat{m}_{k,T}\), the product of the standardized
  induction response rate and the direct conditional maintenance
  estimator above.
\end{itemize}

The primary analysis uses the Horvitz--Thompson form for linear
(risk-difference) durable-response contrasts, since it is
design-unbiased; the direct (Hájek) form \(\hat{m}_{k,T}\) for
conditional maintenance contrasts; and the ratio-normalized pseudo-arm
form on the IPD side of nonlinear (log-odds) contrasts. The comparator
side of those contrasts is reconstructed as the product of published
rates, as described next.

On the comparator side, the same estimand family is reconstructed from
published aggregates: \(q_{k}^{AgD} = r^{AgD} \times m_k^{AgD}\), the
product of the published induction response rate and the published
maintenance response rate among responders re-randomized to arm \(k\).
This requires that the two published rates refer to compatible
populations and denominators, response definitions, rescue rules, and
assessment times. Where the comparator trial reports counts,
\(q_k^{AgD}\) can be computed exactly. Note that \(q_k = r \times m_k\)
is an identity between estimands, not a guarantee about crude published
rates. If the comparator's assignment probabilities varied across
prognostic strata, the crude responder success rate in arm \(k\) need
not estimate maintenance among all induction responders. In that case
the comparator must report an appropriately standardized or
inverse-probability-weighted maintenance estimate, or stratum-specific
counts sufficient to construct one.

In a comparison often described as anchored on withdrawal, the contrast
\(g(q_{jk,T}) - g(q_{j,\mathrm{wd},T})\) is formed within each trial on
a chosen scale \(g\), and the difference between these trial-level
contrasts is then taken. We refer to this as a
\emph{withdrawal-referenced} rather than an anchored comparison, for the
following reason. Since
\(q_{j,\mathrm{wd},T} = r_{j,T} \, m_{j,\mathrm{wd},T}\), each
withdrawal arm inherits the drug-specific induction response rate:
withdrawal after induction with \(A\) and withdrawal after induction
with \(B\) are different sequential treatment strategies. On scales on
which \(r_{j,T}\) does not cancel, such as the log-odds scale (and
likewise the risk-difference scale), the differenced contrast depends on
the difference in induction response rates even when post-withdrawal
maintenance behaviour is identical between the drugs. On the risk-ratio
scale, the induction-response probability cancels within each trial
(\(q_{jk,T}/q_{j,\mathrm{wd},T} = m_{jk,T}/m_{j,\mathrm{wd},T}\)). This
does not change the target population. It only makes the comparison
insensitive to induction response, and it does not relax the
requirements for a valid common comparator. Valid withdrawal-referenced
comparison, on any scale, requires equivalence of the sequential
induction-plus-withdrawal regimens, which is a much stronger condition
than equivalence of withdrawal pharmacology. We return to this in the
Discussion.

Many of these quantities are correlated within each trial: the
standardized induction and durable response rates share calibration
weights and induction data, and in multi-arm trials the contrasts
against a common reference arm are correlated. A two-trial
non-parametric bootstrap accounts for this; each replicate resamples
both evidence sources independently at the level of the individual
participant. The resulting intervals are pointwise, with
participant-level dependencies retained within each evidence source. On
the IPD side, this is a standard non-parametric bootstrap in which each
resampled participant carries all characteristics and outcomes (with
induction non-response coded as maintenance failure), and calibration
(re-weighting) is repeated within each replicate. On the aggregate side,
the same is possible when the publication reports binary outcomes as
complete, mutually exclusive counts, so that every patient who entered
induction can be classified as a non-responder or as a responder
re-randomized to a specific arm. The comparator's outcome-level
pseudo-IPD can then be reconstructed and resampled. Where reporting is
incomplete (e.g.~missing cells or imputed values), additional
assumptions or approximations are required. Comparator resampling
propagates uncertainty in the reconstructed outcome counts while
treating the published baseline calibration targets as fixed. Neither
the uncertainty in those targets nor their covariance with the
comparator outcomes is reconstructed from the available aggregates. The
target of inference is accordingly a population whose baseline moments
equal those reported for the comparator, rather than the comparator's
underlying population.

\hypertarget{regression-adjustment-stc}{%
\subsubsection{Regression adjustment
(STC)}\label{regression-adjustment-stc}}

Regression-based adjustment by simulated treatment comparison
(STC)\textsuperscript{2} follows the same logic but has additional
requirements. The outcome models are identified from the IPD by
factorizing

\[\Pr\{D_i(k) = 1 \mid X_i\} = \Pr(R_i = 1 \mid X_i) \times \Pr\{M_i(k) = 1 \mid R_i = 1, X_i\},\]

where the first factor is estimable from the full induction IPD and the
second from the responders re-randomized to arm \(k\). Restriction to
arm-\(k\) responders is justified for a model conditional on \(X\) when
assignment is independent of the potential maintenance outcomes given
\(X\) and response, as under a fixed ratio. If assignment probabilities
also depend on additional post-induction strata \(S\), the model must
either include those strata or use design weighting, since restriction
to arm \(k\) can otherwise change the distribution of \(S\) at a given
\(X\). The fitted model must then be standardized to the comparator's
induction population, and this depends on that population's joint
covariate distribution. For nonlinear outcome models the joint
distribution is not uniquely identified by marginal means and
proportions. Evaluating the model at the mean covariates is not
equivalent to marginalizing over the population, because correlations,
variances and higher moments matter. STC therefore requires assumptions
about the target joint distribution; for instance, a correlation
structure may be sourced externally or borrowed from the index trial, as
in a previous application\textsuperscript{20}.

\hypertarget{harmonizing-across-trial-designs}{%
\subsubsection{Harmonizing across trial
designs}\label{harmonizing-across-trial-designs}}

RPW allows comparison to be harmonized at the induction-population level
across several study designs. Table 1 summarizes the main
configurations.

\begin{longtable}[]{@{}
  >{\raggedright\arraybackslash}p{(\columnwidth - 6\tabcolsep) * \real{0.2500}}
  >{\raggedright\arraybackslash}p{(\columnwidth - 6\tabcolsep) * \real{0.2500}}
  >{\raggedright\arraybackslash}p{(\columnwidth - 6\tabcolsep) * \real{0.2500}}
  >{\raggedright\arraybackslash}p{(\columnwidth - 6\tabcolsep) * \real{0.2500}}@{}}
\toprule
Index trial (IPD) & Comparator (aggregate) & IPD-side handling & Comparator-side handling \\
\midrule
\endhead
Randomized withdrawal & Randomized withdrawal (any constant ratio) &
Calibrate the full induction cohort to the comparator's induction
aggregates; estimate arm-specific durable response by re-randomization
inverse-probability weighting & \(q_k = r \times m_k\) from published
rates \\
Randomized withdrawal & Treat-through & Calibrate the full induction
cohort; durable response for continued treatment by re-randomization
inverse-probability weighting & Durable response from published
induction response and maintenance among induction responders \\
Treat-through & Randomized withdrawal & Standard induction-population
PAIC; no re-randomization correction (\(D_i = R_i M_i\) observed
directly for all patients on continued treatment) &
\(q_k = r \times m_k\) from published rates \\
Treat-through & Treat-through & Standard PAIC & Standard reporting \\
\bottomrule
\end{longtable}

\emph{Table 1. Design-harmonization configurations. In all cases the
comparison is made at the induction-population level using the
durable-response estimand; the conditional estimand can be obtained as
\(q/r\). The re-randomization ratios of the included trials may differ.}

When the IPD trial uses a treat-through design, durable response is
observed directly and no re-weighting beyond standard PAIC is needed,
but the comparator's aggregates must be recombined to the same estimand.
Treat-through trials also frequently report an overall late-timepoint
response rate in all participants, including those who achieved response
only after the induction cut-off. Randomized-withdrawal trials generally
do not report how many induction non-responders responded later, so this
creates an estimand mismatch. RPW requires harmonization on durable
response, and analysts should verify that the comparator's reported
maintenance rate is conditioned on induction response in the same way as
in the IPD trial.

\hypertarget{suggested-workflow}{%
\subsubsection{Suggested workflow}\label{suggested-workflow}}

The analysis proceeds roughly as follows:

\begin{enumerate}
\def\labelenumi{\arabic{enumi}.}
\tightlist
\item
  Confirm that response definitions, assessment times, rescue rules, and
  analysis sets are compatible across trials.
\item
  Specify the target population, and the effect modifiers and prognostic
  factors to adjust for.
\item
  Estimate calibration weights \(a_i\) on the full induction cohort;
  check convergence, weighted balance, and effective sample size.
\item
  Estimate \(\hat{r}_T\), \(\hat{q}_{k,T}\) and \(\hat{m}_{k,T}\) using
  the known \(\pi_{ik}\) and the estimator forms of the preceding
  section.
\item
  Reconstruct the comparator's estimand family from its published
  counts.
\item
  Estimate intervals with the two-trial bootstrap.
\item
  Report effective sample sizes and balance diagnostics by arm, and
  conduct sensitivity analyses.
\end{enumerate}

\hypertarget{synthetic-demonstration}{%
\subsection{Synthetic demonstration}\label{synthetic-demonstration}}

We demonstrate RPW using three synthetic trials modelled on commonly
used designs, in which responders to a 16-week induction period are
re-randomized to continued treatment at different dosing intervals or to
withdrawal. All data are synthetic and represent no actual drug, trial,
or patient. Because the data-generating models are known, the true value
of every estimand can be computed and the behaviour of the estimators
verified. All three synthetic trials use fixed re-randomization ratios,
so the fixed-ratio weights apply exactly. The complete R code, which
reproduces every number and figure in this section from fixed seeds, is
provided in a stand-alone R file available on
zenodo.org\textsuperscript{24}.

We simulate four baseline covariates: age, sex, baseline severity, and
prior systemic therapy. Severity and age are prognostic for response at
both induction and maintenance, prior systemic therapy is additionally
prognostic for induction response, and severity also acts as an effect
modifier of induction response. The covariate distributions differ
between the three trials, so naïve comparison is confounded.

\begin{itemize}
\tightlist
\item
  \textbf{Trial A, index with IPD} (Drug A, \(n = 800\)): randomized
  withdrawal; week-16 responders re-randomized 2:2:1 to high dose, low
  dose, and withdrawal (\(\pi = 0.4, 0.4, 0.2\)). Non-responders are not
  followed beyond induction.
\item
  \textbf{Trial B} (Drug B, aggregate data only, \(n = 700\)):
  randomized withdrawal; week-16 responders re-randomized 2:1 to high
  dose or withdrawal. The publication reports baseline summaries for the
  induction population, the induction response rate, and maintenance
  response in each arm among induction responders, but not baseline
  characteristics of the responder subset.
\item
  \textbf{Trial C} (Drug C, aggregate data only, \(n = 650\)):
  treat-through, with high-dose treatment to week 52. The publication
  reports induction-population baseline summaries, week-16 response,
  week-52 response among week-16 responders, and overall week-52
  response including late responders.
\end{itemize}

Each trial could also have included an initial randomization to placebo;
this is immaterial here, as a placebo arm would not be used.

Figure 1 illustrates the three designs. Trial C's published overall
week-52 response (0.362) exceeds its durable response rate (0.274)
because it includes late responders, an illustrative instance of the
estimand mismatch noted above.

\begin{figure}
\centering
\includegraphics{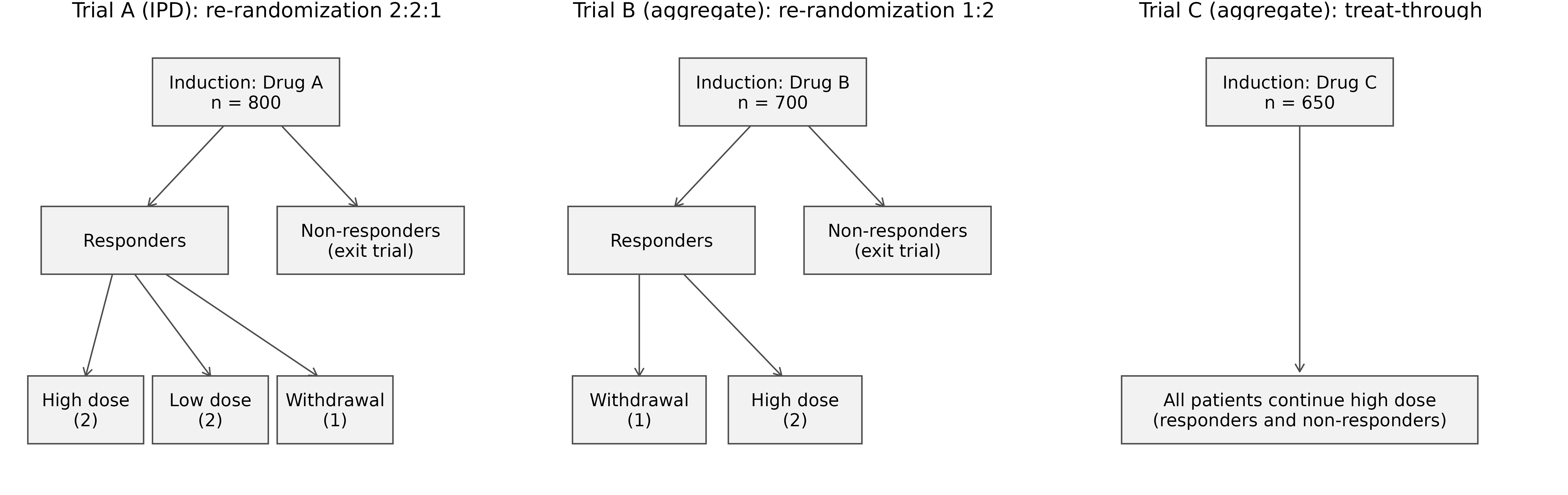}
\caption{The three synthetic trial designs: re-randomization 2:2:1
(Trial A), re-randomization 1:2 (Trial B), treat-through (Trial C).}
\end{figure}

\hypertarget{baseline-reconstruction}{%
\subsubsection{Baseline reconstruction}\label{baseline-reconstruction}}

Table 2 and Figure 2 display properties of Trial A. Induction responders
are a visibly selected subset: their mean baseline severity is 26.1,
against 28.2 in the full induction population, and mean age and the
proportion with prior systemic therapy also differ. Applying the RPW
design weights approximates the induction-population summaries within
each pseudo-arm in this realization.

\begin{longtable}[]{@{}lllll@{}}
\toprule
Population & Age & Severity & Male & Prior systemic \\
\midrule
\endhead
Induction population (\(n = 800\)) & 38.3 & 28.2 & 0.55 & 0.41 \\
Responders only & 37.1 & 26.1 & 0.55 & 0.37 \\
Pseudo-arm: high dose & 38.0 & 28.3 & 0.56 & 0.40 \\
Pseudo-arm: low dose & 38.6 & 28.4 & 0.56 & 0.43 \\
Pseudo-arm: withdrawal & 38.1 & 27.6 & 0.50 & 0.41 \\
\bottomrule
\end{longtable}

\emph{Table 2. Weighted baseline means in Trial A. Each RPW pseudo-arm
(responders assigned to the arm, weight 1, plus all non-responders,
weight \(\pi_k\)) approximates the induction-population means up to
sampling variation; the smaller withdrawal arm
(\(\pi_{\mathrm{wd}} = 0.2\)) shows the largest residual noise. Realized
counts in this realization: 347 responders of \(n = 800\), re-randomized
152 (high dose), 122 (low dose) and 73 (withdrawal).}

\begin{figure}
\centering
\includegraphics{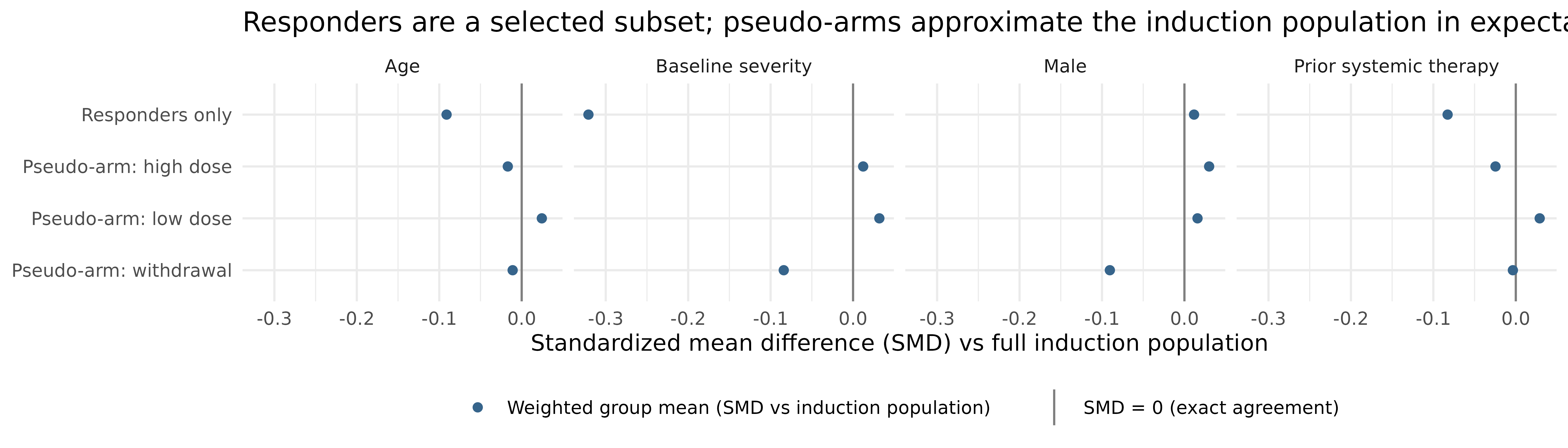}
\caption{Standardized mean differences versus the full induction
population: responders are selected; the pseudo-arms approximate the
induction population in expectation. A single small pseudo-arm can
remain imbalanced in a given realization, as the withdrawal arm's sex
difference shows.}
\end{figure}

\hypertarget{indirect-comparisons}{%
\subsubsection{Indirect comparisons}\label{indirect-comparisons}}

Using only the Trial A IPD and the published aggregates of Trials B and
C, we conduct three comparisons, each both naively and with adjustment:

\begin{enumerate}
\def\labelenumi{\arabic{enumi}.}
\tightlist
\item
  an unanchored MAIC of A versus B on durable response and on
  conditional maintenance under high-dose continuation, re-weighting
  Trial A to Trial B's induction population;
\item
  a withdrawal-referenced comparison of A versus B on the log-odds scale
  of durable response, to illustrate its precision cost and, as shown
  below, that it is not a comparison against a common anchor; and
\item
  an unanchored MAIC of A versus C (treat-through) on durable response.
\end{enumerate}

MAIC weights match the means of all four covariates plus the second
moment of severity. The second-moment target is computed from the
published mean \(\bar{x}\) and standard deviation \(s\) on the
comparator's own denominator, as \(\bar{x}^2 + (n-1)s^2/n\). All
contrasts are A minus comparator. The adjusted analysis is labelled RPW
+ MAIC. The reconstructed pseudo-population serves as the balance
diagnostic of Table 2, and estimation uses only the weights described in
Methods. Confidence intervals are pointwise 95\% percentile intervals
from a two-trial bootstrap with 1,000 replicates. Each replicate
resamples the Trial A IPD at the individual level, with calibration
weights re-estimated, and independently resamples the reconstructed
outcome-level pseudo-data of the comparator. Because the world is
synthetic, each comparison's true value in its target population is
known and is reported alongside the estimates.

\begin{longtable}[]{@{}
  >{\raggedright\arraybackslash}p{(\columnwidth - 6\tabcolsep) * \real{0.2500}}
  >{\raggedright\arraybackslash}p{(\columnwidth - 6\tabcolsep) * \real{0.2500}}
  >{\raggedright\arraybackslash}p{(\columnwidth - 6\tabcolsep) * \real{0.2500}}
  >{\raggedright\arraybackslash}p{(\columnwidth - 6\tabcolsep) * \real{0.2500}}@{}}
\toprule
Comparison & Truth & Naive (95\% CI) & RPW + MAIC (95\% CI) \\
\midrule
\endhead
A vs B durable response, unanchored (risk difference) & 0.053 & 0.157
(0.090 to 0.226) & 0.084 (0.014 to 0.156) \\
A vs B conditional maintenance, unanchored (risk difference) & 0.068 &
0.024 (−0.072 to 0.125) & −0.003 (−0.115 to 0.105) \\
A vs B durable response, withdrawal-referenced (log-OR difference) &
0.130 & 0.313 (−0.352 to 1.062) & 0.503 (−0.206 to 1.317) \\
A vs C durable response, unanchored (risk difference) & 0.039 & 0.092
(0.027 to 0.164) & 0.049 (−0.013 to 0.118) \\
\bottomrule
\end{longtable}

\emph{Table 3. Primary analysis of one simulated realization, with
pointwise 95\% two-sided bootstrap intervals for both methods. The naive
analysis ignores population differences; the adjusted analysis applies
RPW design weights and MAIC calibration weights. The
withdrawal-referenced row is what would typically be called
``anchored'', with the withdrawal arms used as an anchor, assumed
equivalent. Effective sample sizes of the combined
calibration-and-design weights,
\(\mathrm{ESS}(w) = (\sum_i w_i)^2 / \sum_i w_i^2\), after calibration
to Trial B: 465 (induction population), 342 (high-dose pseudo-arm), 277
(withdrawal pseudo-arm), from \(n = 800\) in Trial A. Comparator outcome
cells (Trial B, \(n = 700\)): 196 responders, of whom 66 were
re-randomized to withdrawal (19 maintained response) and 130 to high
dose (97 maintained response).}

\begin{figure}
\centering
\includegraphics{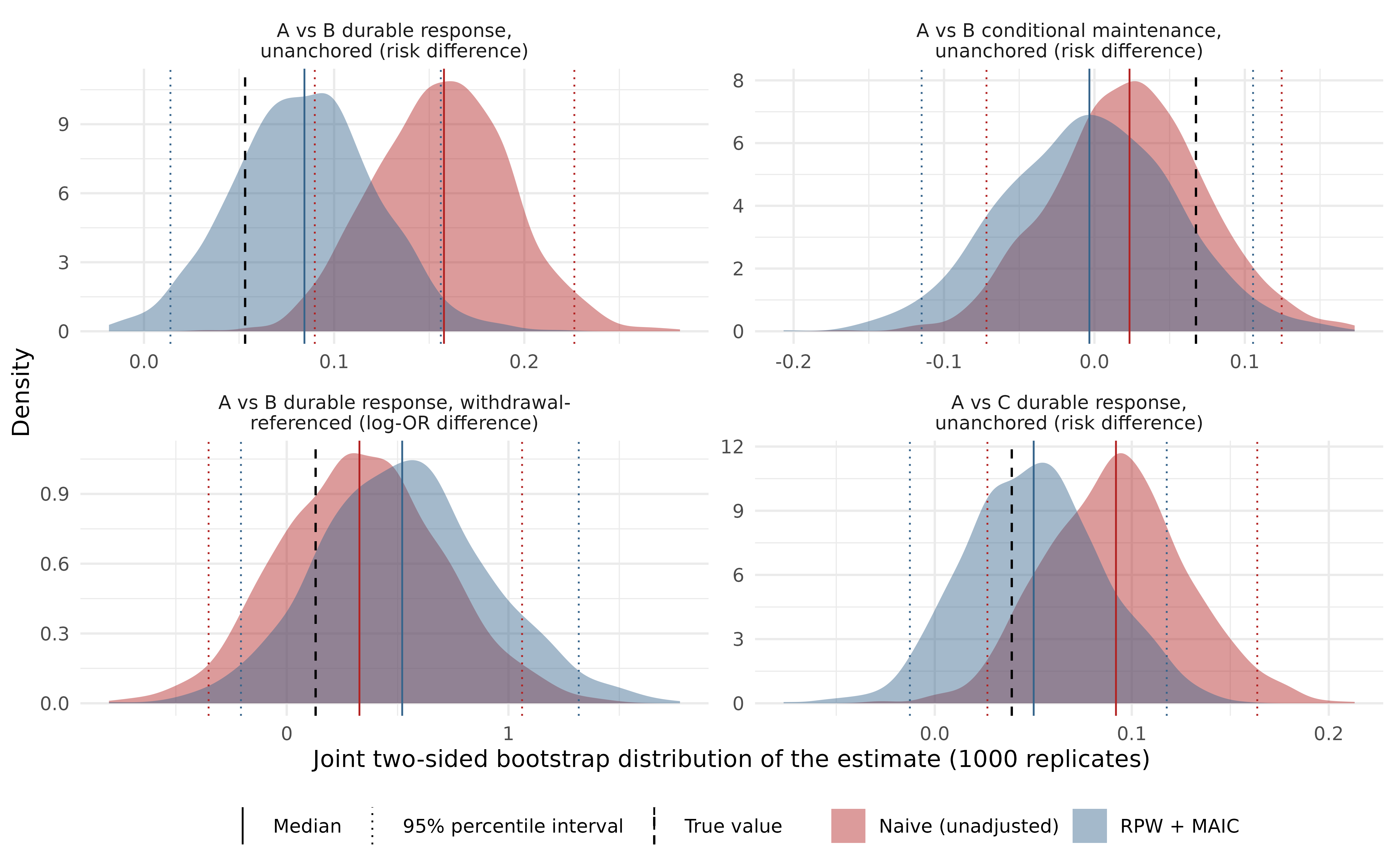}
\caption{Joint two-sided bootstrap distributions of every primary
estimate (naive in red, adjusted in blue; semi-transparent densities),
with vertical lines marking each distribution's median (solid), 2.5th
and 97.5th percentiles (dotted), and the true value (dashed black). The
withdrawal-referenced (``anchored'') contrast is displayed on the log-OR
scale, where its bootstrap distribution is approximately normal; on the
odds-ratio scale the corresponding distribution is approximately
log-normal.}
\end{figure}

In this realization, adjustment substantially improves the point
estimates of the unanchored durable-response comparisons (A vs B: 0.157
naive, 0.084 adjusted, truth 0.053; A vs C: 0.092 naive, 0.049 adjusted,
truth 0.039). For A versus B, the naive interval (0.090 to 0.226)
excludes the true value. The adjusted conditional-maintenance comparison
is −0.003 against a truth of 0.068, with an interval (−0.115 to 0.105)
that contains the truth. The repeated-sampling results below show that
this estimator is centred correctly, so the discrepancy is merely an
unfortunate draw. The withdrawal-referenced comparison is noisy, resting
on the small withdrawal arms of both trials.

The demonstration also illustrates the problem with anchoring on
withdrawal described in Methods. The withdrawal maintenance model is
identical for the two drugs by construction. The true conditional
maintenance probabilities under withdrawal, standardized to Trial B's
population, are 0.308 for drug A and 0.304 for drug B. However, the true
withdrawal-referenced log-OR contrast is 0.130, of which 0.025 arises
purely because the two drugs' induction response rates differ (0.309
versus 0.264). Recomputing the contrast with drug B's induction response
set equal to drug A's, holding all maintenance behaviour fixed, gives
0.105. Equivalence of the withdrawal condition therefore does not make
the withdrawal arms a common anchor on the durable-response scale. On
the risk-ratio scale each regimen's induction-response probability
cancels within its own trial (the true rate-free contrast is 0.079).
That contrast remains an induction-population quantity, algebraically
equal to the responder-conditional one. It compares relative
continuation-versus-withdrawal effects across regimens rather than the
continued treatments directly. This decomposition is specific to the
scenario and the effect scale; it illustrates the mechanism and is not a
general measure of the induction-rate component.

\hypertarget{operating-characteristics}{%
\subsubsection{Operating
characteristics}\label{operating-characteristics}}

Figure 4 reports a point-estimator experiment in which Trial A is
regenerated 2,000 times against fixed population-level comparator
aggregates, isolating the expectation of the estimators. Naive mean bias
is +0.106 for the unanchored A-versus-B durable-response risk difference
and +0.055 for A versus C; after RPW + MAIC it is +0.001 in both cases.
For the conditional-maintenance risk difference, bias is +0.035 naive
and +0.002 after RPW + MAIC. The adjusted withdrawal-referenced log-OR
contrast shows a small plug-in bias (+0.04, against a replicate standard
deviation of roughly 0.33). This is expected for smooth nonlinear
functionals of re-randomized arms, even in their ratio-normalized
(Hájek-type) form. In this scenario the naive withdrawal-referenced
estimator happens to show less mean bias (+0.007), because referencing
to withdrawal removes much of the population confounding here and the
remaining effect-modifier imbalance is modest. This is a property of the
simulated scenario, not of the estimators. Both observations are
reported as observed: adjustment did not improve every estimator in this
scenario.

\begin{figure}
\centering
\includegraphics{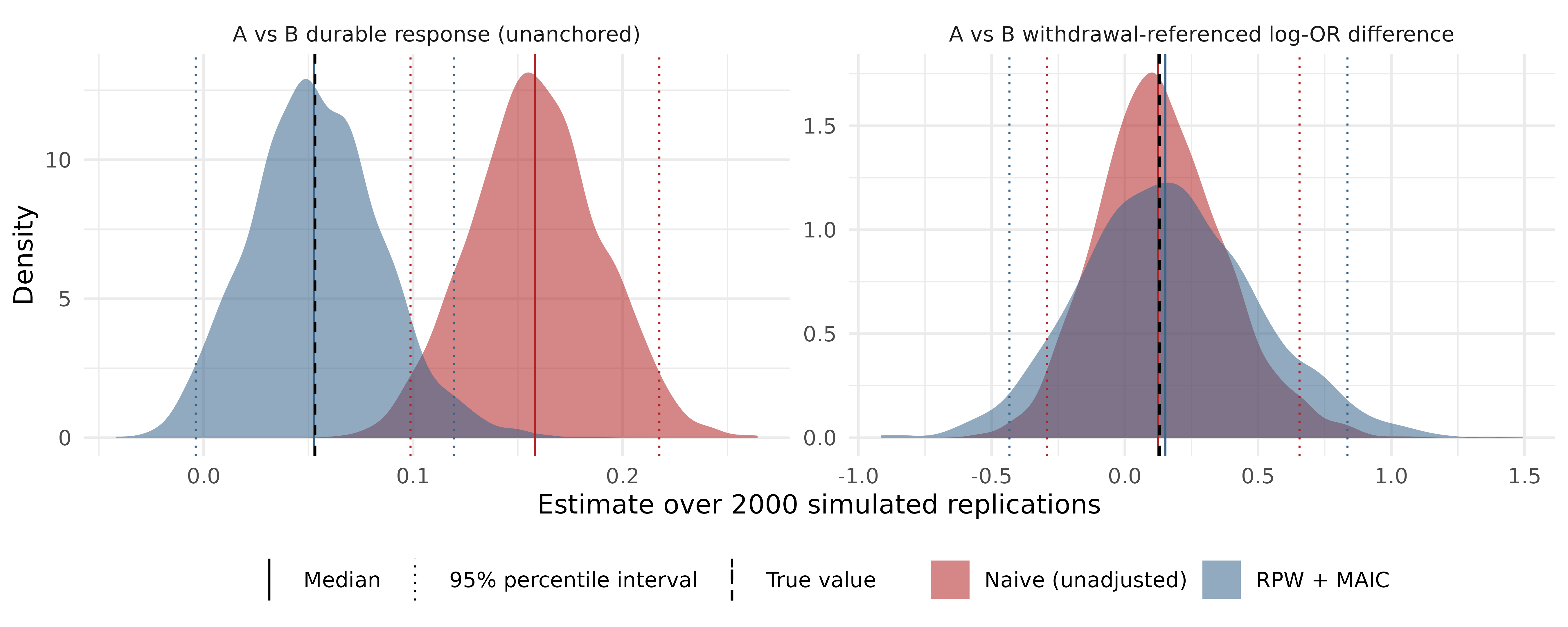}
\caption{Sampling distributions over 2,000 replications: the naive
estimator of the unanchored durable-response contrast is biased, while
the RPW + MAIC estimator is centred on the truth. Vertical lines mark
each distribution's median (solid), 2.5th and 97.5th percentiles
(dotted), and the true value (dashed black).}
\end{figure}

The full procedure, including interval construction, was evaluated
separately. In each of 400 replications, fresh finite versions of all
three trials were drawn, Trials B and C were published, and the
two-trial bootstrap was run with 200 resamples. Table 4 shows the
operating characteristics of the adjusted estimators.

\begin{longtable}[]{@{}
  >{\raggedright\arraybackslash}p{(\columnwidth - 10\tabcolsep) * \real{0.1667}}
  >{\raggedright\arraybackslash}p{(\columnwidth - 10\tabcolsep) * \real{0.1667}}
  >{\raggedright\arraybackslash}p{(\columnwidth - 10\tabcolsep) * \real{0.1667}}
  >{\raggedright\arraybackslash}p{(\columnwidth - 10\tabcolsep) * \real{0.1667}}
  >{\raggedright\arraybackslash}p{(\columnwidth - 10\tabcolsep) * \real{0.1667}}
  >{\raggedright\arraybackslash}p{(\columnwidth - 10\tabcolsep) * \real{0.1667}}@{}}
\toprule
Estimand (adjusted) & Bias & Empirical SE & RMSE & Mean bootstrap SE & Coverage \\
\midrule
\endhead
A vs B durable response (risk difference) & +0.003 & 0.035 & 0.035 &
0.034 & 94.3\% \\
A vs B conditional maintenance (risk difference) & +0.007 & 0.063 &
0.063 & 0.064 & 94.3\% \\
A vs B withdrawal-referenced (log-OR) & +0.046 & 0.388 & 0.390 & 0.399 &
91.8\% \\
A vs C durable response (risk difference) & +0.002 & 0.034 & 0.034 &
0.033 & 91.5\% \\
\bottomrule
\end{longtable}

\emph{Table 4. Operating characteristics of the adjusted estimators over
400 full-procedure replications: bias, empirical standard error, root
mean squared error, mean bootstrap standard error, and coverage of the
95\% percentile interval. Monte Carlo standard errors of the reported
coverages are approximately 1.2 percentage points at 94.3\% and 1.4
percentage points at 91.8\% and 91.5\%.}

The mean bootstrap standard errors track the empirical standard errors
closely. Coverage is near nominal for the two unanchored A-versus-B
contrasts but lower for the withdrawal-referenced and A-versus-C
contrasts (91.8\% and 91.5\%). With Monte Carlo standard errors of
approximately 1.2 to 1.4 percentage points at 400 replications, this
undercoverage is modest but real. Its cause has not been isolated.
Plausible contributors are plug-in bias, skewness on the log-odds scale,
the ordinary finite-sample behaviour of percentile intervals, and Monte
Carlo error in tail quantiles estimated from only 200 inner bootstrap
draws. Bootstrap replicates with non-finite log-odds contrasts are
excluded when computing percentile intervals, and the analysis code
reports their frequency.

We ran three stress scenarios, each over 1,000 replications against
population-level comparator aggregates (Table 5). First, stratified
assignment: re-randomization probabilities in Trial A differ by a
baseline-severity stratum (2:2:1 below severity 30, 1:1:1 at or above).
The correct analysis with patient-specific \(W_{ik}\) is compared with
the incorrect application of the overall marginal share as a single
fixed fraction for all patients. Second, an omitted effect modifier:
calibration of Trial A to Trial B matches the means of age, sex, and
prior systemic therapy but omits severity, which is designed to be both
prognostic and an effect modifier. Third, correlated covariates: age and
severity are correlated 0.5 in the target population but uncorrelated in
the index trial, so the joint distributions differ in a way that
matching on marginal moments cannot remove.

\begin{longtable}[]{@{}
  >{\raggedright\arraybackslash}p{(\columnwidth - 6\tabcolsep) * \real{0.2500}}
  >{\raggedright\arraybackslash}p{(\columnwidth - 6\tabcolsep) * \real{0.2500}}
  >{\raggedright\arraybackslash}p{(\columnwidth - 6\tabcolsep) * \real{0.2500}}
  >{\raggedright\arraybackslash}p{(\columnwidth - 6\tabcolsep) * \real{0.2500}}@{}}
\toprule
Scenario (A vs B durable response, risk difference) & Bias & Empirical SE & RMSE \\
\midrule
\endhead
Stratified assignment, correct patient-specific \(W_{ik}\) & +0.001 &
0.032 & 0.032 \\
Stratified assignment, marginal fraction applied to all (incorrect) &
−0.006 & 0.030 & 0.031 \\
Omitted effect modifier (severity unmatched) & +0.073 & 0.034 & 0.081 \\
Correlated covariates, marginal moment matching & −0.005 & 0.030 &
0.031 \\
\bottomrule
\end{longtable}

\emph{Table 5. Stress scenarios for the unanchored durable-response
contrast (A vs B, risk difference), 1,000 replications each. The correct
stratified analysis remains centred; the marginal-fraction shortcut and
unmatched joint structure produce small but systematic bias in this
configuration, and omitting an effect modifier produces severe bias, as
the identification conditions predict.}

The correct stratified analysis remains centred, whereas the
marginal-fraction shortcut is systematically biased. Omitting an effect
modifier from the calibration produces bias an order of magnitude
larger; RPW does not relax any of the requirements of a properly
conducted PAIC. Finally, the inability of MAIC on marginal moments to
reflect differing correlations between adjustment factors leaves a small
residual bias.

\hypertarget{discussion}{%
\subsection{Discussion}\label{discussion}}

This article presents a simple, design-based procedure that permits PAIC
of binary maintenance outcomes where one or more of the included trials
use responder re-randomization. Re-randomization-probability weighting
harmonizes comparison for each maintenance arm by leveraging weights
matching the trial's re-randomization probabilities, as a randomized
sample of the full induction population, whose baseline characteristics
are typically well reported. Failure coding of maintenance in induction
non-responders defines a durable-response outcome on the induction
population, such that standard MAIC or STC methods can be used. Trials
employing different re-randomization ratios, and randomized-withdrawal
vs.~treat-through designs, can be harmonized in whichever direction IPD
is available.

\hypertarget{relation-to-prior-work}{%
\subsubsection{Relation to prior work}\label{relation-to-prior-work}}

The arithmetic used in RPW has previously been exploited in NMA settings
in a more limited fashion. In the NMA literature, re-randomized
maintenance trials have been converted to reflect approximate
treat-through arms by use of ``dilution'' of induction non-responders by
the re-randomization factor\textsuperscript{9}. This dilution approach
has later been used in several disease areas\textsuperscript{10--12}.
The dilution conversion matches the aggregate-count special case of the
weights presented here, i.e.~dividing the non-responder count (and event
count) by three is the same as applying weight \(\pi_k = 1/3\) to each
non-responder. The equivalence concerns the weighting arithmetic only:
in that conversion the diluted non-responder events (there, 129 events
among induction non-responders, diluted to 43) remain in the maintenance
numerator, so the converted arm targets a late-timepoint response
outcome, whereas the durable-response endpoint used here fixes the
outcome contribution of induction non-responders at zero. Furthermore,
the inverse-probability representation of the same identity is standard
in sequentially randomized (SMART) designs\textsuperscript{21,22}, and
decomposition of a treat-through response probability into induction
response times conditional maintenance has been used explicitly in
applied MAICs\textsuperscript{16}. The contribution here is the explicit
application and operationalization of this established design weighting
for induction-level population adjustment under a reporting constraint:
the same identity holds for arbitrary functions of baseline covariates
(e.g.~means, proportions, raw moments) and can therefore be used where
baseline characteristics have not been reported for the maintenance
population. Some precursors have applied known re-randomization
proportions to patient and event counts only\textsuperscript{9,11}, with
HTA reviewers criticizing such an applied conversion for assuming that
responders and non-responders have the same baseline
characteristics\textsuperscript{13}. That criticism concerns the applied
outcome-level conversion and subsequent comparison; it is not a
condition of the within-trial weighting identity, which permits
systematic baseline differences between responders and non-responders.
Others have reconstructed the comparator's maintenance-phase baseline
means as responder-proportion-weighted averages of induction cohort
means\textsuperscript{19}. While conceptually similar, weighting
cohort-level means by responder prevalence does not itself reconstruct
within-cohort responder covariate means, and the approach carries no
design-based guarantee. In cases where baseline characteristics have
been disclosed for the responder sub-population, applied MAICs,
including our own earlier comparison of maintenance outcomes in atopic
dermatitis\textsuperscript{17}, have simply matched on
them\textsuperscript{16,18}, which is fine under specific assumptions
regarding the comparability of maintenance populations.

Principal stratification\textsuperscript{25} addresses a related but
separate problem: causal comparison within a latent stratum defined by
the joint potential response statuses under the treatments being
compared (for example, patients who would respond under both regimens).
It targets the post-treatment-selection estimand problem head-on, at the
cost of substantially stronger identifying assumptions, and is generally
impractical in the one-IPD-trial-versus-aggregates setting of PAIC. RPW
instead sidesteps responder-level comparison by reformulating the target
at the induction-population level.

RPW has already been applied: my team has used the approach in several
indirect-comparison projects over the last five years, including several
used in HTA submissions. An STC application, including the use of
pairwise non-parametric bootstrapping for uncertainty estimates, is the
first to have reached open publication\textsuperscript{20}. This paper
is intended as the method's reference description: providing in-depth
formal presentation, giving the general inverse-probability form, its
conditions, its estimators, and its limits. This should accordingly be
read as a further development and generalization of a conceptual
approach found in the NMA dilution literature, developed independently
of that literature and connected to it after the fact.

\hypertarget{alternative-responder-level-comparison}{%
\subsubsection{Alternative responder-level
comparison}\label{alternative-responder-level-comparison}}

In cases where baseline characteristics for comparator re-randomized
arms have been disclosed, an alternative responder-level comparison is
available: match the index trial's responders directly to the reported
maintenance-baseline characteristics and compare maintenance outcomes at
responder level. This has been done in applied maintenance
MAICs\textsuperscript{16,18}, and relies on a critical assumption that
the maintenance outcome among induction responders does not depend on
which drug produced the response during induction, conditional on the
matched maintenance-baseline covariates. This assumption should be
stated explicitly\textsuperscript{17}. Matching on characteristics at
re-randomization is not obviously weaker than matching on pre-treatment
baselines, and could arguably capture more of the responder selection.
Anything not measured that separates responder populations cannot be
balanced, including depth or mechanism of response beyond the matched
scores and residual drug exposure entering the maintenance phase. Direct
responder matching and induction-population standardization generally
target different responder distributions: under induction-population
standardization each regimen retains its own response-selection
mechanism, whereas responder matching averages the index outcomes over
the comparator's responder population. Where both are feasible, their
results may be compared descriptively, but disagreement need not
indicate bias; a formal sensitivity comparison requires an explicitly
shared estimand and additional assumptions linking the responder
populations. Within the hierarchy presented below, this route falls into
the unanchored category with a different conditioning set.

\hypertarget{anchored-unanchored-or-neither}{%
\subsubsection{Anchored, unanchored, or
neither?}\label{anchored-unanchored-or-neither}}

RPW technically supports both anchored and unanchored comparison, but
they do not make the same requirements. The essential problems with
using withdrawal arms as an anchor are conceptual and structural, not
pharmacological. As shown in Methods,
\(q_{j,\mathrm{wd},T} = r_{j,T} \, m_{j,\mathrm{wd},T}\) includes the
response rate following the induction treatment. Withdrawal following
induction with \(A\) and withdrawal following induction with \(B\) are
different sequential strategies, not a shared comparator condition. On
the log-odds scale of durable response the induction rates do not
cancel, which in our simulation demonstration resulted in roughly one
fifth of the true withdrawal-referenced contrast being attributable to
differences in induction response rates alone, despite withdrawal
behaviour being identical by construction. On the risk-ratio scale the
shared induction-response probability cancels within each regimen.
However, the contrast remains the induction-population risk ratio for
durable response under the two sequential strategies, which is
algebraically identical to a maintenance risk ratio among that regimen's
responders. The cancellation renders the comparison insensitive to
induction-response prevalence. It does, however, not magically make
responder populations generated by different induction treatments
comparable, nor does it turn the withdrawal strategies into a common
comparator. The cross-trial ratio of risk ratios compares relative
continuation-versus-withdrawal effects across induction regimens, not
continued treatments directly. Valid withdrawal-referenced comparison
therefore theoretically requires equivalence of the full
induction-plus-withdrawal regimens on the chosen effect scale, not just
equivalent withdrawal pharmacology.

Equivalence in the withdrawal condition itself, necessary though not
sufficient, is also demanding: it fails whenever the offset dynamics of
the two drugs differ: pharmacokinetic persistence, rebound phenomena,
tapering schedules, rescue and escape rules, and assessment timing all
impact what a ``withdrawal'' arm actually experiences. Drug persistence
alone can span a large fraction of the maintenance period: in
SOLO-CONTINUE, median dupilumab serum concentrations fell below the
limit of quantification only by week 12 or week 24 after
re-randomization to placebo, depending on the prior dose
regimen\textsuperscript{7}. While the drug was not observable at the end
of the maintenance period, variable persistence may impact the available
period in which loss of response could occur. Drugs sharing modality,
target, and half-life can still differ in post-withdrawal durability of
response through mechanisms such as differences in dissociation
kinetics, binding affinity, or persistence of downstream pharmacodynamic
effects; these are candidate explanations to be argued in an
application, not established facts of any given comparison. The
published withdrawal arms may provide a partial testing opportunity: if
the reconstructed withdrawal-arm durable-response rates differ notably
between trials after population adjustment, equivalence is arguably
untenable. Unfortunately, observed near-equivalence alone is
insufficient to establish actual equivalence. The evidence required to
support equivalence of sequential regimens is accordingly strong, and I
suspect that obtaining it may be difficult or unfeasible in the settings
wherein PAIC are typically applied.

This suggests the following hierarchy for PAIC of maintenance outcomes.
The best option is to apply a conventional common anchor in cases where
this is available (for example, a placebo arm maintained in parallel
through the maintenance period in both trials). Where a common anchor is
not available, use unanchored comparison of durable response where the
demanding requirements of unanchored comparison (all prognostic factors
and effect modifiers accounted for\textsuperscript{4}) are judged
credible. Treat withdrawal-referenced contrasts as potential
supplementary sensitivity analyses. This can be defended only under
explicitly argued assumptions regarding equivalent standardized
induction-plus-withdrawal regimens. Finally, accept that in some
real-life cases, none of these conditions can be defended, and the
available evidence simply fails to support a reliable comparative
estimate. Stating this is arguably preferable to presenting a comparison
based on assumptions that cannot be justified.

The withdrawal arms are not useless, however. RPW applies to every
maintenance arm, so the sequential strategies ``induce with \(j\), then
withdraw responders'' allows for unanchored comparison. Such
strategy-level comparisons have already been used, including
interpolation on the log-odds scale between the maintenance and
withdrawal strategies to estimate comparative durability under varying
levels of treatment maintenance, the ``durability
index''\textsuperscript{20}; interpreting interpolated values as
outcomes under actual partial adherence rests on additional assumptions
beyond the design identity. At the induction-population level,
\(q_{j,\mathrm{wd},T}\) compares the treatment followed by withdrawal
strategies as policies, and answers a relevant clinical question: of
patients administered induction treatment with each drug, what
proportion achieves response and retains it after stopping treatment? At
responder level, \(m_{j,\mathrm{wd},T}\) addresses another relevant
clinical question: to what extent can response be expected to survive
stopping treatment?

\hypertarget{limitations}{%
\subsubsection{Limitations}\label{limitations}}

The method restores the baseline covariate distribution in randomization
expectation, and allows PAIC under the usual modeling assumptions; it
does not, by itself, guarantee an unbiased treatment comparison.
Specifically:

\begin{enumerate}
\def\labelenumi{\arabic{enumi}.}
\tightlist
\item
  \textbf{Expectation, not realization.} Equality of covariate
  distributions holds in expectation over the re-randomization
  mechanism; realized maintenance arms are subject to Monte Carlo
  variation. Weighted balance and arm-specific ESS must be reported.
\item
  \textbf{All standard PAIC requirements remain.} These conditions are
  outside the scope of this paper but are not relaxed by it.
\item
  \textbf{Endpoint and missing-data conventions.} Failure-coding of
  induction non-responders is an estimand definition, not an imputation,
  and does not justify failure-coding of maintenance-phase dropouts
  unless that rule is part of the endpoint in both trials.
\item
  \textbf{Binary durable-response outcomes only.} The estimand is
  defined through the binary gate \(R_i\): \(D_i = R_i M_i\) and is
  unambiguous because lack of response following induction fixes the
  outcome at zero. There is no general equivalent for continuous
  maintenance endpoints.
\item
  \textbf{Post-treatment selection for the conditional estimand}, as
  discussed above.
\item
  \textbf{Withdrawal arms are not a general anchor,} as discussed in
  detail under ``Anchored, unanchored, or neither?''.
\item
  \textbf{Design fidelity.} Assignment probabilities that differ across
  patients or strata require the patient- or stratum-specific
  \(\pi_{ik}\) in the inverse-probability form \(W_{ik}\); the
  fractional pseudo-re-randomization weights are valid only when a
  common assignment probability applies to all patients, and a single
  marginal probability applied where probabilities vary can invalidate
  the reconstruction. Stratified randomization applying the same ratio
  within every stratum retains a common probability.
\item
  \textbf{Nonlinear functionals.} Design-unbiasedness applies to
  weighted covariate moments and linear functionals; contrasts on
  nonlinear scales (odds ratios, log-odds) are subject to ordinary
  small-sample plug-in bias even when the ratio-normalized (Hájek-type)
  estimator forms are used, as demonstrated in our simulation.
\end{enumerate}

\hypertarget{conclusions}{%
\subsubsection{Conclusions}\label{conclusions}}

Re-randomization-probability weighting recreates, in expectation, the
induction population's baseline characteristics within each maintenance
arm of a randomized-withdrawal trial, employing nothing beyond the
design parameters of the trials. Combined with failure-coding of
induction non-responders, it allows population-adjusted indirect
comparisons of binary maintenance outcomes between trials using
different re-randomization ratios and between randomized-withdrawal and
treat-through designs, comparisons that are otherwise at the mercy of
the empirically spotty reporting on the baseline characteristics of
re-randomized responders. Where no valid common anchor is available and
the assumptions required for unanchored population adjustment are
credible, durable response provides an induction-population estimand for
the primary comparison; otherwise, the available evidence may not
support a reliable comparative estimate. Withdrawal-referenced contrasts
require equivalence of sequential induction-plus-withdrawal regimens and
should be treated as sensitivity analyses. Re-randomization-probability
weighting is transparent, fully reproducible from published design
information, and makes existing PAIC methods applicable in specific
cases where they previously were not.

\hypertarget{declarations}{%
\subsection{Declarations}\label{declarations}}

\textbf{Funding:} No specific funding was received for this work.

\textbf{Conflicts of interest:} In addition to his position as Senior
Researcher at Akershus University Hospital, Lørenskog, Norway, Kim Rand
is a partner and principal at Maths in Health B.V., a consultancy
conducting health economics and outcomes research for clients in the
pharmaceutical industry, academia, and the public sector. Rand has
conducted a large number of ITC projects and other consultancy work for
various clients in the pharmaceutical industry. Neither the development
of these methods nor the manuscript have received support from any
clients.

\textbf{Author contributions (CRediT):} Kim Rand: conceptualization,
methodology, software, formal analysis, writing (original draft, review
and editing).

\textbf{Ethics approval:} Not applicable. The included demonstration
uses synthetic data and published aggregate information; no human
participants were involved.

\textbf{Data and code availability:} The manuscript uses no real patient
data. The complete R script \texttt{appendix\_A\_simulation.R} is
available on zenodo.org\textsuperscript{24} (base R plus ggplot2, fixed
seeds isolated per stage) reproduces every number, table and figure in
the Synthetic demonstration section when run top-to-bottom from its own
directory: the three synthetic trials, the baseline-reconstruction
diagnostic (Table 2, Figure 2), the unanchored and withdrawal-referenced
analyses with the two-trial bootstrap (Table 3, Figure 3), the
2,000-replication point-estimator experiment (Figure 4), the coverage
study and stress scenarios (Tables 4 and 5), the standardized population
quantities quoted in the anchoring-decomposition paragraph (the
withdrawal maintenance probabilities, the induction response rates, and
the counterfactual and risk-ratio contrasts, printed in the script's
section 6b), and unit checks of the design-weight identities. It writes
\texttt{simulation\_summary.csv},
\texttt{operating\_characteristics\_B.csv},
\texttt{operating\_characteristics\_stress.csv} and
\texttt{sessionInfo.txt}, and saves each figure as both PNG (600 dpi,
white background) and PDF. The primary demonstration uses fixed
re-randomization ratios, for which the fixed-ratio weights apply
exactly; the stratified scenario exercises the general per-patient form.

\textbf{Use of AI:} Claude Opus v4.8 (Anthropic; accessed through Claude
Code) was used to edit the text in early August 2026, including prompts
to identify internal inconsistencies and incomplete rationales and to
suggest wording changes improving legibility. Claude Opus was also used
to tidy up, improve documentation of, and stress test the demonstration
code. Claude Fable v5 (Anthropic; accessed through Claude Code) was used
in September 2026 to review internal consistency. All methodological
content, analyses, and conclusions are the author's own.

\hypertarget{references}{%
\subsection{References}\label{references}}

\hypertarget{refs}{}
\begin{CSLReferences}{0}{0}
\leavevmode\vadjust pre{\hypertarget{ref-signorovitch2010}{}}%
\CSLLeftMargin{1. }
\CSLRightInline{Signorovitch JE, Wu EQ, Yu AP, et al. Comparative
effectiveness without head-to-head trials: a method for
matching-adjusted indirect comparisons applied to psoriasis treatment
with adalimumab or etanercept. PharmacoEconomics. 2010;28(10):935-945.
Published online 2010.
doi:\href{https://doi.org/10.2165/11538370-000000000-00000}{10.2165/11538370-000000000-00000}}

\leavevmode\vadjust pre{\hypertarget{ref-caro2010}{}}%
\CSLLeftMargin{2. }
\CSLRightInline{Caro JJ, Ishak KJ. No Head-to-Head Trial? Simulate the
Missing Arms. PharmacoEconomics. 2010;28(10):957-967. Published online
2010.
doi:\href{https://doi.org/10.2165/11537420-000000000-00000}{10.2165/11537420-000000000-00000}}

\leavevmode\vadjust pre{\hypertarget{ref-phillippo2018}{}}%
\CSLLeftMargin{3. }
\CSLRightInline{Phillippo DM, Ades AE, Dias S, Palmer S, Abrams KR,
Welton NJ. Methods for Population-Adjusted Indirect Comparisons in
Health Technology Appraisal. Med Decis Making. 2018;38(2):200-211.
Published online 2018.
doi:\href{https://doi.org/10.1177/0272989X17725740}{10.1177/0272989X17725740}}

\leavevmode\vadjust pre{\hypertarget{ref-phillippo2016_guidance}{}}%
\CSLLeftMargin{4. }
\CSLRightInline{Phillippo DM, Ades AE, Dias S, Palmer S, Abrams KR,
Welton NJ. NICE DSU Technical Support Document 18: Methods for
Population-Adjusted Indirect Comparisons in Submissions to NICE.
Sheffield: NICE Decision Support Unit; 2016. Published online 2016.}

\leavevmode\vadjust pre{\hypertarget{ref-signorovitch2012}{}}%
\CSLLeftMargin{5. }
\CSLRightInline{Signorovitch JE, Sikirica V, Erder MH, et al.
Matching-Adjusted Indirect Comparisons: A New Tool for Timely
Comparative Effectiveness Research. Value Health. 2012;15(6):940-947.
Published online 2012.
doi:\href{https://doi.org/10.1016/j.jval.2012.05.004}{10.1016/j.jval.2012.05.004}}

\leavevmode\vadjust pre{\hypertarget{ref-wollenberg2021_baselines}{}}%
\CSLLeftMargin{6. }
\CSLRightInline{Wollenberg A, Blauvelt A, Guttman-Yassky E, et al.;
ECZTRA 1 and ECZTRA 2 study investigators. Tralokinumab for
moderate-to-severe atopic dermatitis: results from two 52-week,
randomized, double-blind, multicentre, placebo-controlled phase III
trials (ECZTRA 1 and ECZTRA 2). Br J Dermatol. 2021;184(3):437-449.
Published online 2021.
doi:\href{https://doi.org/10.1111/bjd.19574}{10.1111/bjd.19574}}

\leavevmode\vadjust pre{\hypertarget{ref-worm2020}{}}%
\CSLLeftMargin{7. }
\CSLRightInline{Worm M, Simpson EL, Thaci D, et al. Efficacy and Safety
of Multiple Dupilumab Dose Regimens After Initial Successful Treatment
in Patients With Atopic Dermatitis: A Randomized Clinical Trial. JAMA
Dermatol. 2020;156(2):131-143. Published online 2020.
doi:\href{https://doi.org/10.1001/jamadermatol.2019.3617}{10.1001/jamadermatol.2019.3617}}

\leavevmode\vadjust pre{\hypertarget{ref-blauvelt2023_baselines}{}}%
\CSLLeftMargin{8. }
\CSLRightInline{Blauvelt A, Thyssen JP, Guttman-Yassky E, et al.
Efficacy and safety of lebrikizumab in moderate-to-severe atopic
dermatitis: 52-week results of two randomized double-blinded
placebo-controlled phase III trials. Br J Dermatol. 2023;188(6):740-748.
Published online 2023.
doi:\href{https://doi.org/10.1093/bjd/ljad022}{10.1093/bjd/ljad022}}

\leavevmode\vadjust pre{\hypertarget{ref-thorlund2015}{}}%
\CSLLeftMargin{9. }
\CSLRightInline{Thorlund K, Druyts E, Toor K, Jansen JP, Mills EJ.
Incorporating alternative design clinical trials in network
meta-analyses. Clin Epidemiol. 2015;7:29-35. Published online 2015.
doi:\href{https://doi.org/10.2147/CLEP.S70853}{10.2147/CLEP.S70853}}

\leavevmode\vadjust pre{\hypertarget{ref-thorlund2015b}{}}%
\CSLLeftMargin{10. }
\CSLRightInline{Thorlund K, Druyts E, Toor K, Mills EJ. Comparative
efficacy of golimumab, infliximab, and adalimumab for moderately to
severely active ulcerative colitis: a network meta-analysis accounting
for differences in trial designs. Expert Rev Gastroenterol Hepatol.
2015;9(5):693-700. Published online 2015.
doi:\href{https://doi.org/10.1586/17474124.2015.1024657}{10.1586/17474124.2015.1024657}}

\leavevmode\vadjust pre{\hypertarget{ref-welty2020}{}}%
\CSLLeftMargin{11. }
\CSLRightInline{Welty M, Mesana L, Padhiar A, Naessens D, Diels J, van
Sanden S, Pacou M. Efficacy of ustekinumab vs. advanced therapies for
the treatment of moderately to severely active ulcerative colitis: a
systematic review and network meta-analysis. Curr Med Res Opin.
2020;36(4):595-606. Published online 2020.
doi:\href{https://doi.org/10.1080/03007995.2020.1716701}{10.1080/03007995.2020.1716701}}

\leavevmode\vadjust pre{\hypertarget{ref-disher2025}{}}%
\CSLLeftMargin{12. }
\CSLRightInline{Disher T, Naessens D, Sanon M, Bonner A, Ellis J,
Bartlett M, Hooper B, Yang Z, Allegretti JR, Dignass A. One-Year
Efficacy of Guselkumab Versus Advanced Therapies for the Treatment of
Moderately to Severely Active Crohn's Disease: A Network Meta-Analysis.
Adv Ther. 2025;42(6):2708-2727. Published online 2025.
doi:\href{https://doi.org/10.1007/s12325-025-03183-x}{10.1007/s12325-025-03183-x}}

\leavevmode\vadjust pre{\hypertarget{ref-nice2022}{}}%
\CSLLeftMargin{13. }
\CSLRightInline{National Institute for Health and Care Excellence.
Ozanimod for treating moderately to severely active ulcerative colitis
{[}TA828{]}: committee papers (evidence review group report). London:
NICE; 2022. Published online 2022.}

\leavevmode\vadjust pre{\hypertarget{ref-panaccione2023}{}}%
\CSLLeftMargin{14. }
\CSLRightInline{Panaccione R, Collins EB, Melmed GY, et al. Efficacy and
Safety of Advanced Therapies for Moderately to Severely Active
Ulcerative Colitis at Induction and Maintenance: An Indirect Treatment
Comparison Using Bayesian Network Meta-analysis. Crohns Colitis 360.
2023;5(2):otad009. Published online 2023.
doi:\href{https://doi.org/10.1093/crocol/otad009}{10.1093/crocol/otad009}}

\leavevmode\vadjust pre{\hypertarget{ref-barberio2025}{}}%
\CSLLeftMargin{15. }
\CSLRightInline{Barberio B, Gracie DJ, Black CJ, Ford AC. Network
Meta-Analysis: Efficacy of Biological Therapies and Small Molecules as
Maintenance Therapy in Ulcerative Colitis. Aliment Pharmacol Ther.
2025;62(1):4-21. Published online 2025.
doi:\href{https://doi.org/10.1111/apt.70209}{10.1111/apt.70209}}

\leavevmode\vadjust pre{\hypertarget{ref-reinisch2024}{}}%
\CSLLeftMargin{16. }
\CSLRightInline{Reinisch W, Melmed GY, Nakase H, et al. Comparative
Efficacy and Safety of Upadacitinib vs. Vedolizumab, Ustekinumab, and
Tofacitinib After Induction and Maintenance for Ulcerative Colitis:
Three Matching-Adjusted Indirect Comparisons. Adv Ther.
2024;41(10):3832-3849. Published online 2024.
doi:\href{https://doi.org/10.1007/s12325-024-02912-y}{10.1007/s12325-024-02912-y}}

\leavevmode\vadjust pre{\hypertarget{ref-rand2024}{}}%
\CSLLeftMargin{17. }
\CSLRightInline{Rand K, Ramos-Goni JM, Akmaz B, Sole-Feu L, Armario-Hita
JC. Matching-Adjusted Indirect Comparison of the Long-Term Efficacy
Maintenance of Lebrikizumab and Dupilumab in Patients with
Moderate-to-Severe Atopic Dermatitis. Dermatol Ther (Heidelb).
2024;14(1):169-182. Published online 2024.
doi:\href{https://doi.org/10.1007/s13555-023-01058-z}{10.1007/s13555-023-01058-z}}

\leavevmode\vadjust pre{\hypertarget{ref-augustin2025}{}}%
\CSLLeftMargin{18. }
\CSLRightInline{Augustin M, Armstrong A, Issa NT, et al. Anchored
Matching-Adjusted Indirect Comparison of the Long-Term Maintenance of
Efficacy of Tralokinumab and Lebrikizumab in Patients with
Moderate-to-Severe Atopic Dermatitis. Dermatol Ther (Heidelb).
2026;16(1):277-291. Published online 2026.
doi:\href{https://doi.org/10.1007/s13555-025-01563-3}{10.1007/s13555-025-01563-3}}

\leavevmode\vadjust pre{\hypertarget{ref-jairath2025}{}}%
\CSLLeftMargin{19. }
\CSLRightInline{Jairath V, Raine T, Leahy TP, et al. Matching-adjusted
indirect comparisons of efficacy outcomes between etrasimod and ozanimod
for moderately to severely active ulcerative colitis. J Comp Eff Res.
2025;14(4):e240193. Published online 2025.
doi:\href{https://doi.org/10.57264/cer-2024-0193}{10.57264/cer-2024-0193}}

\leavevmode\vadjust pre{\hypertarget{ref-silverberg2026}{}}%
\CSLLeftMargin{20. }
\CSLRightInline{Silverberg JI, Irvine AD, Foley P, et al. A Durability
Index for Atopic Dermatitis: Indirect Comparison of Lebrikizumab,
Dupilumab, and Tralokinumab in Maintaining Efficacy Under Variable
Treatment Adherence. Dermatol Ther (Heidelb). 2026. Published online
2026.
doi:\href{https://doi.org/10.1007/s13555-026-01791-1}{10.1007/s13555-026-01791-1}}

\leavevmode\vadjust pre{\hypertarget{ref-murphy2005}{}}%
\CSLLeftMargin{21. }
\CSLRightInline{Murphy SA. An experimental design for the development of
adaptive treatment strategies. Stat Med. 2005;24(10):1455-1481.
Published online 2005.
doi:\href{https://doi.org/10.1002/sim.2022}{10.1002/sim.2022}}

\leavevmode\vadjust pre{\hypertarget{ref-nahumshani2012}{}}%
\CSLLeftMargin{22. }
\CSLRightInline{Nahum-Shani I, Qian M, Almirall D, et al. Experimental
design and primary data analysis methods for comparing adaptive
interventions. Psychol Methods. 2012;17(4):457-477. Published online
2012. doi:\href{https://doi.org/10.1037/a0029372}{10.1037/a0029372}}

\leavevmode\vadjust pre{\hypertarget{ref-horvitz1952}{}}%
\CSLLeftMargin{23. }
\CSLRightInline{Horvitz DG, Thompson DJ. A Generalization of Sampling
Without Replacement From a Finite Universe. J Am Stat Assoc.
1952;47(260):663-685. Published online 1952.
doi:\href{https://doi.org/10.1080/01621459.1952.10483446}{10.1080/01621459.1952.10483446}}

\leavevmode\vadjust pre{\hypertarget{ref-rand2026_code}{}}%
\CSLLeftMargin{24. }
\CSLRightInline{Rand K.
intelligentaccident/rerandomization\_probability\_weighting: R code
accompanying paper on re-randomization probability weighting for use in
maintenance PAIC where trials include responder re-randomization.
Version v1.0.0 {[}Software{]}. Zenodo; 2026. Published online 2026.
doi:\href{https://doi.org/10.5281/zenodo.23240397}{10.5281/zenodo.23240397}}

\leavevmode\vadjust pre{\hypertarget{ref-frangakis2002}{}}%
\CSLLeftMargin{25. }
\CSLRightInline{Frangakis CE, Rubin DB. Principal Stratification in
Causal Inference. Biometrics. 2002;58(1):21-29. Published online 2002.
doi:\href{https://doi.org/10.1111/j.0006-341X.2002.00021.x}{10.1111/j.0006-341X.2002.00021.x}}

\end{CSLReferences}

\end{document}